\documentclass[letterpaper
    ,final            
  ]
  {aipproc}

\layoutstyle{6x9}

\usepackage[T1]{fontenc}
\usepackage[latin1]{inputenc}
\usepackage{amsmath}
\usepackage{amssymb}

\usepackage{bm}
\usepackage{slashed}
\usepackage{braket}
\usepackage{nicefrac}

\newcommand{\wt}[1]{\widetilde{#1}}

\newcommand{\PVTV}{\slashed{P}\slashed{T}}

\begin{document}

\title{Schiff Theorem and the Electric Dipole Moments of Hydrogen-Like Atoms}

\classification{11.30.Er, 24.80.+y, 32.80.Ys}
\keywords      {Schiff theorem, Schiff moment, electric dipole moment, time-reversal violation}

\author{C.-P. Liu}{
  address={Kernfysisch Versneller Instituut, Zernikelaan 25, 9747 AA Groningen, The Netherlands},
  altaddress={Theoretical Division, Los Alamos National Laboratory, Los Alamos, NM 87545, USA}
}

\author{W. C. Haxton}{
  address={INT and Dept. of Physics, Univ. of Washington, Box 351550, Seattle, WA 98195-1550, USA}
}

\author{M. J. Ramsey-Musolf}{
  address={Kellogg Radiation Laboratory, California Institute of Technology, Pasadena, CA 91125, USA}
}

\author{R. G. E. Timmermans}{
  address={Kernfysisch Versneller Instituut, Zernikelaan 25, 9747 AA Groningen, The Netherlands}
}

\author{A. E. L. Dieperink}{
  address={Kernfysisch Versneller Instituut, Zernikelaan 25, 9747 AA Groningen, The Netherlands}
}

\begin{abstract}
The Schiff theorem is revisited in this work and the residual $P$-
and $T$-odd electron--nucleus interaction, after the shielding takes
effect, is completely specified. An application is made to the electric dipole moments of hydrogen-like
atoms, whose qualitative features and systematics have important implication for 
realistic paramagnetic atoms. 
\end{abstract}

\maketitle


\section{Introduction}

The permanent electric dipole moment (EDM) of a physical system is
an indication of time-reversal ($T$) violation which, by $CPT$ invariance,
is equivalent to $CP$ violation, one of the most profound puzzles in 
elementary particle physics. Although a neutral atom is ideal 
for such a precision measurement, much of its EDM evades detection 
because of the re-arrangement of its constituents in order to screen
the applied electric field and keep the whole system stationary.

As the shielding is not exact, many experiments have been
performed over the years to measure these tiny, residual EDMs with
gradually improved techniques and accuracy. Since an atom contains
both electrons and nucleons, its EDM receives contributions from all
possible $P$- and $T$-odd ($\PVTV$) dynamics in leptonic, semi-leptonic,
and hadronic sectors. The first part of this work is to completely specify, 
after incorporating the shielding effect, the residual $\PVTV$ electron--nucleus ($e\,N$) interaction, $\wt{H}_{e\,N}$. The second part concerns 
an application to the EDMs of hydrogen-like (H-like) atoms where some general features and 
systematics of the contributions from different $\PVTV$ sources are extracted.



\section{Schiff Theorem Revisited}

The so-called Schiff theorem states the following: for a nonrelativistic 
system made up of point, charged particles which interact electrostatically
with each other and with an arbitrary external field, the shielding
is complete~\cite{Schiff:1963}. Applying this theorem to atoms,
the assumptions of this theorem are not exactly satisfied because: (1) the
atomic electrons can be quite relativistic, (2) the
atomic nucleus has a finite structure, and (3) the electromagnetic
(EM) interaction between $e\,e$ or $e\,N$ has magnetic components. Therefore, a measured atomic
EDM, $d_{A}$, or any upper bound on it, is a combined manifestation
of these effects. This can be best summarized by $\wt{H}_{e\,N}$,
through which $d_{A}$ can be expressed as 
\begin{align}
d_{A}\equiv\langle\bm d_{A}\rangle & \cong\sum_{n}\,\frac{1}{E_{0}-E_{n}}\,\left(\bra{0}e\,\bm x\ket{n}\bra{n}\wt{H}_{e\,N}\ket{0}+\textrm{c.c.}\right)\,,\label{eq:d_A}
\end{align}
where $\ket{0}$ and $\ket{n}$ represent the atomic ground and excited
states, respectively.

While the implementation of the shielding effect,
which can be carried out by various ways 
\citep[see, {\it e.g.},][]{Schiff:1963,Sandars:1968b,Feinberg:1977,Sushkov:1984,Engel:1999np,Sandars:2001nq,Flambaum:2001gq}
is too tedious to be shown here, a simplified example below can be used to
illustrate the basic points. The key relationship is to re-write the
internal EDM interaction $\wt{H}_{e\,N}^{d}$ as a commutator
involving the unperturbed atomic Hamiltonian, $H_{0}$, plus some
remaining terms, if any, 
\begin{align}
\wt{H}_{e\,N}^{d} & =[\bm d\cdot\bm\nabla\,,\, H_{0}]\,+\ldots\,.\label{eq:shielding}\end{align}
Substituting the commutator in Eq.~(\ref{eq:d_A}), a closure sum
can be performed and leads to the result: $-\langle\bm d\rangle$.
Therefore, if there is no term left beyond the commutator, $\wt{H}_{e\,N}^{d}$
then contributes to the total EDM exactly opposite to $\langle\bm d\rangle$:
this is the complete shielding.

It should be emphasized that the Schiff theorem is a quantum-mechanical
description of the shielding effect, and this implies that Eq.~(\ref{eq:shielding})
should be realized at the operator level, \emph{i.e.}, every quantity
is operator. While we obtained the similar expression for the residual
interaction due to the electron EDM, $d_{e}$, as Refs.~\cite{Sandars:1968b,Lindroth:1989a},
the residual interaction due to the nuclear EDM, $d_{N}$,
differs from existing literature. These differences can be summarized
in the Schiff moment, $S$, we found 
\begin{align}
S \equiv \langle \bm S \rangle &  = \frac{e}{10}\,\left(\langle r^{2}\,\bm r\rangle-\frac{5}{3}\,\frac{1}{Z}\,\left\langle \left[r^{2}\,(1-\frac{4\,\sqrt{\pi}}{5}\, Y_{2}(\hat{r}))\otimes\bm r\right]_{1}\right\rangle \right)\,,\label{eq:Schiff moment new}\end{align}
in contrast to the usual definition \citep[see, {\it e.g., }][]{Sushkov:1984,Engel:1999np,Sandars:2001nq,Flambaum:2001gq}. 

Besides the additional quadrupole term we include in the derivation
(it was usually ignored), the main difference is a matrix element
of the composite operator $\langle r^{2}\otimes\bm r\rangle$ in the former
versus a product of the matrix elements $\langle r^{2}\rangle\ \otimes \langle\bm r\rangle$
in the latter, where "$\otimes$" denotes the recoupling of angular momenta. This stems from the way we treat $\bm d_{N}\propto\bm r$
as an operator in Eq.~(\ref{eq:shielding}), and the previous definition
already takes the matrix element $\langle\bm d_{N}\rangle\propto\langle\bm r\rangle$
before the derivation. As the nucleons in the nucleus do not necessarily
respond in a coherent manner after the shielding sets in, our definition
accounts for these additional dynamics which is left
out in the previous one. A calculation for deuteron shows quite some difference:
in $S$, the three terms contribute as $1:-5/3:-4/3$; in
the traditional definition, supposed the quadrupole term is included as $\left[\langle r^{2}\, Y_{2}\rangle\otimes\langle\bm r\rangle\right]_{1}$, the ratio becomes $1:-0.59:-0.07$. While the huge difference
in deuteron can be attributed to its loose binding, other nuclear Schiff
moments should be revised, because they receive most contributions
from the surface region, where the binding is usually not as strong
as in the core. 

Our derivation also treated the finite-size effect more carefully,
following the suggestion of  Ref.~\cite{Flambaum:2001gq}, and included all the magnetic
$e\,N$ interactions into account. As a result, the full
form of $\wt{H}_{e\,N}$ contains all possible nuclear moments,
either long-ranged or local, charge or magnetic. The details will
be presented in a later publication~\cite{Liu:2005c}.

\section{Electric Dipole Moments of Hydrogen-Like Atoms}



The parity
admixture of an atomic ground state, $\wt{\ket{0}}$ ($1s_{1/2}$
for H-like atoms) can be solved from the Sternheimer equation~\cite{Sternheimer:1954}. For the cases where the Pauli approximation
is valid, the results can be expressed analytically. Separating the contributions
from (1) $d_{e}$, (2) $C_{\textrm{PS,S}}^{0}$, a representative
case in the semi-leptonic $e\,N$ interaction,
which is isoscalar and nuclear spin independent, (3) $S$, and (4)
$S^{\textrm{mag}}$, the magnetic equivalence of $S$,
which contains, \emph{e.g}., the magnetic quadrupole moment, they
roughly grow with the atomic number $Z$ as $Z^{2}$, $Z\, A$, $Z\,S$,
and $Z\,S^{\textrm{mag}}$. The common factor $Z^{1}$ comes
from the atomic structure calculation of $\wt{\ket{0}}$, and the remaining
growth factor indicates how the corresponding $\PVTV$ interaction
scales. As $S$ and $S^{\textrm{mag}}$ both involve
$r^{2}$-weighted nuclear moments, they roughly scale with $A^{2/3}$.

The main reason that heavy paramagnetic atoms are suitable for constraining
$d_{e}$ is usually justified from a $Z^{3}$ enhancement factor~\citep[see, {\it e.g.}, ][]{Sandars:1966,Flambaum:1976vg}.
Based on the systematics found in H-like atoms, actually part of the
enhancement, $Z^{2}$, comes with $\wt{\ket{0}}$, and can be crudely
understood from the fact that it takes less energy for a $p$-state
excitation from a $ns_{1/2}$ state with $n > 1$ than
from $1s_{1/2}$. As this $Z^{2}$ enhancement
also applies to other $\PVTV$ sources, the competition between the
four contributors mentioned above only goes with $Z$, $A$, $S$, and $S^{\textrm{mag}}$.
Suppose the semi-leptonic or hadronic $\PVTV$ intreactions are large,
\emph{i.e.}, large $C$ or $S$ and $S^{\textrm{mag}}$,
then the dominance of the $d_{e}$ contribution is questionable. In
this sense, it is better to have a series of EDM measurements and
use their results to constrain these $\PVTV$ sources simultaneously,
or at least, one should semi-quantitatively determine the conditions under which
$d_{e}$ is clearly the winner. In either case, the calculations of
semi-leptonic and hadronic contributions are indispensable for a more
thorough study of EDMs of paramagnetic atoms.





\bibliographystyle{aipproc}   


\bibliography{PANIC05}

\end{document}